\documentclass[epj]{svjour}
\usepackage{graphics}
\begin{document}
%
%

%
\title{A collision-induced satellite in the Lyman~$\beta$ profile
          due to H-H collisions}
\subtitle{Lyman~$\beta$ satellites}
\author{N.F. Allard \inst{1} 
\and J. Kielkopf \inst{2} 
\and I. Drira \inst{3}
\and P. Schmelcher \inst{4}
}                     
\offprints{N.F. Allard}          
\institute{
      Observatoire de Paris-Meudon, D\'epartement Atomes et
      Mol\'ecules en Astrophysique, 92195 Meudon Principal Cedex, 
      France \\ 
      CNRS Institut d'Astrophysique, 98 bis Boulevard Arago, 75014
      Paris, France  \\
      email: allard@iap.fr
\and 
      Department of Physics, University of Louisville,
      Louisville, Ky 40292 U.S.A. \\
      email: john@aurora.physics.louisville.edu   
\and
      Observatoire de Paris-Meudon, D\'epartement Atomes et
      Mol\'ecules en Astrophysique, 92195 Meudon Principal Cedex, 
      France \\
      email:Ikhlas.Drira@obspm.fr
\and
      Institut f\"{u}r Physikalische Chemie, Universit\"{a}t Heidelberg,
      Im Neuenheimer Feld 229, 69120 Heidelberg, Germany \\
      email: peter@tc.pci.uni-heidelberg.de
                                              }
\date{Received: date / Revised version: date}
%
\abstract{
We present a theoretical profile of the 
Lyman~$\beta$  line of atomic hydrogen perturbed by collisions 
with neutral hydrogen atoms and  protons.
We use a general unified theory in which the electric dipole moment
varies during a collision. 
A collision-induced satellite appears on Lyman~$\beta$, correlated to the 
$\mathrm{B}''\bar{\mathrm{B}}\;^1\Sigma^+_u - \mathrm{X}\;^1\Sigma^+_g$ 
asymptotically forbidden transition of H$_2$.  
As a consequence,
the appearance of the
line wing between Lyman~$\alpha$ and  Lyman~$\beta$ is shown to be
sensitive to the relative abundance of hydrogen ions and neutral atoms,
and thereby to provide a temperature diagnostic for
stellar atmospheres and laboratory plasmas.
} 
\maketitle
\section{Introduction}
\label{intro}
In Allard {\it et al.}\/ 1999~\cite{allard99} we 
derived a classical path expression for a pressure-broadened atomic
spectral line shape that allows for a radiative electric dipole transition
moment which 
depends on the position of the  perturbers. This factor  is not included in the
more usual approximation of Anderson \& Talman 1956~\cite{anderson}
and Baranger 1958~\cite{baranger58a,baranger58b}. 
We used this theory to  study the influence of the
variation of the dipole moment on the satellites present in the far
wing profiles of the Lyman series lines of atomic hydrogen seen in stars and
in laboratory plasmas.

Satellite features at 1600~\AA\/  and 1405~\AA\/  in
the Lyman~$\alpha$ wing associated with free-free  quasi-molecular
transitions of H$_2$ and H$_2^+$ have been observed in
ultraviolet (UV) spectra of  certain stars  obtained with the
International Ultraviolet Explorer (IUE) and the
Hubble Space Telescope 
(HST)~\cite{koester93,koester94,holweger94,bergeron95}. 
The stars which show Lyman~$\alpha$ satellites are 
DA white dwarfs, old {\it Horizontal} {\it Branch}\/ stars of spectral
type A, and the $\lambda$~Bootis stars. The last two have the
distinctive  property of  poor {\it metal}\/ content, that is, low
abundances  of elements other than H and He.  
Satellites also have been observed in the laboratory spectra of
laser-produced hydrogen  plasmas~\cite{kielkopf95,kielkopf98}.

Satellite features in hydrogen lines are not limited to Lyman~$\alpha$, which
is  the only Lyman-series line accessible to IUE as well as HST.  
Observations 
with HUT (Hopkins Ultraviolet Telescope)
and  with ORFEUS (Orbiting Retrievable Far and Extreme Ultraviolet
Spectrograph)
of 
Lyman~$\beta$ of DA white dwarfs with
T$_{\rm eff}$ close to 20000~K have revealed a line shape 
very different from the
expected  simple Stark broadening, with line satellites near 1078 and
1060~\AA\/ \cite{koester96,koester98}.
The satellites in the red wing of Lyman~$\beta$ are in the 905
 to 1187~\AA\/ spectral region covered by the Far Ultraviolet Spectroscopic
Explorer (FUSE) launched in June 1999. Furthermore,
Lyman~$\beta$ profiles are also the subject of an ongoing study of
the far ultraviolet spectrum of dense hydrogen plasmas.
The strengths of these satellite features and indeed the entire shape
 of  wings in the  Lyman series are very sensitive to the  degree of
 ionization in the stellar atmosphere and laboratory plasmas, because that
determines the relative importance of broadening by ion and neutral
collisions.

In a previous work~\cite{allard98a} we 
presented  theoretical profiles of Lyman~$\beta$ perturbed  solely by
protons. The calculations were based on the
accurate theoretical H$_2^+$ molecular potentials of
Madsen and Peek~\cite{madsen71} 
to describe the interaction between radiator and perturber,  
and dipole transition moments of
Ramaker and Peek~\cite{ramaker72}. The line profiles were included
as a source of opacity  in model atmospheres for hot white dwarfs, 
and the predicted spectra compared very well with the observed
ORFEUS  spectra \cite{koester98}. 

The very recent {\it ab initio}\/ calculations of Drira 1999~\cite{drira99}
of electronic transition
moments for excited states of the H$_2$ molecule
and very accurate molecular potentials of Schmelcher 
\cite{schmelcher,detmer98}
now allow us to compute Lyman~$\beta$ profiles simultaneously perturbed
by neutral atomic hydrogen and by protons.
The aim of our paper is to  point out a collision-induced satellite
correlated to the 
$\mathrm{B}''\bar{B}\;^1\Sigma^+_u - \mathrm{X}\;^1\Sigma^+_g$
asymptotically forbidden transition of H$_2$.
We show  that the shape of the wing in
the region between Lyman~$\beta$ and Lyman~$\alpha$ is
particularly sensitive  
to the relative abundance of the neutral and ion perturbers responsible
for the broadening of the lines.

\section{Theory}

The classical path theory for the shape of 
pressure-broadened atomic spectral lines which takes into account the variation
of the electric dipole moment during a collision  is only briefly 
outlined here. The  theory has been described in detail in~\cite{allard99}.
Our approach is based on the quantum theory of
spectral line shapes of Baranger~\cite{baranger58a,baranger58b}  developed in
an  {\it adiabatic representation}\/ to include the degeneracy of atomic
levels~\cite{royer74,royer80,allard94}.

\subsection{General expression for the spectrum in an adiabatic
 representation }

\label{sec:theory}
The spectrum $I(\omega)$ can be written as the Fourier transform
 of the dipole autocorrelation function $\Phi$(s),
\begin{equation}
I(\omega)=\frac{1}{\pi} \, Re \, \int^{+\infty}_0\Phi(s)e^{-i \omega s} ds \; .
\end{equation}
Here, 
\begin{eqnarray}
\Phi(s)
& = & {\bf Tr} \, \rho  {\bf D}^{\dagger}
e^{ \frac{ is{\bf H} }{ \hbar} }
{\bf D} e^{    \frac{-is{\bf H} }{ \hbar}   } \\
& = & \langle {\bf D}^{\dagger}(0) 
{\bf D}(s) \rangle  \label{eq:autocorr} 
\end{eqnarray}
is the autocorrelation function of \mbox{ ${\bf D}(s)$ },
the dipole moment of the radiator in the Heisenberg 
representation (we use bold notation for operators)~\cite{allard82}.
${\bf H}$ is the total Hamiltonian 
\begin{equation}
{\bf H} = {\bf T}_{nucl} + {\bf T}_{elec} +  V({\bf x},{\bf R}),
\end{equation} 
where  ${\bf T}_{nucl}$ and ${\bf T}_{elec}$ are sums of nuclear and electronic
kinetic energy operators respectively, and $V({\bf x},{\bf R})$ is the
interaction between particles.
Here ${\bf x}$ denotes collectively the set of electronic coordinates
(position and spin) plus spin coordinates of the nuclei, while ${\bf R}$
denotes the set of position coordinates of the nuclei. 
We assume that the radiating atom is immersed in a
perturber bath in thermal equilibrium.
The density matrix $\rho$ is
\begin{equation}
\rho \equiv \frac{          e^{-\beta {\bf H}  }    }
                 { {\bf Tr} \, e^{-\beta {\bf H}  }    } \; ,
\end{equation}
where $\beta$ is the inverse temperature ($1/kT$).

We use the notation 
\begin{equation}
\langle (\; ) \rangle  \equiv  {\bf Tr} \, \rho (\; ) \; ,
\end{equation}
where $ {\bf Tr} $ denotes the trace operation.

The adiabatic or Born-Oppenheimer representation
comprises expanding states of the gas  in terms of electronic states
$\chi_e(x;R)$ corresponding to frozen nuclear configurations.
In the Schr\"{o}dinger equation
\begin{eqnarray}
( {\bf T}_{elec} +  V({\bf x},{\bf R}) ) \chi_e(x;R) &=&  
{\bf H}_{elec}(R)\chi_e(x;R) \\
&=& E_e(R) \chi_e(x;R) \; .
\end{eqnarray}
$R$ appears as a parameter, and the
eigenenergies $E_e(R)$ play the role of   potential energies for the
nuclei.
Any total wave function $\Psi(x,R)$ can be
expanded as 
\begin{equation}
\Psi(x,R) = \sum_e \psi_e(R) \, \chi_e(x;R) \; .
\end{equation}

As the nuclei get far from each other, which we denote by 
\mbox{ $R \rightarrow \infty$ },
the electronic energies $E_e(R)$ tend to asymptotic values 
$E_e^{\infty}$ which are sums of individual atomic energies. 
Since atomic states
are usually degenerate, there are in general several different
energy surfaces which tend to a same asymptotic energy as 
\mbox{ $R \rightarrow \infty$ }.
 We will consider specifically a single radiating atom, the 
{\em radiator},  immersed in a gas of optically inactive atoms, 
the {\em perturbers}.
For a transition $\alpha =(i,f)$ from
initial state $i$ to final state $f$, we have 
$R$-dependent  frequencies
\begin{eqnarray}
\omega_{e'e}(R) \equiv (E_{e'}(R)-E_e(R))/\hbar \; , &   \; \;  \; \; e  \; \in
\varepsilon_i  \; ,  \;  e'  \; \in \varepsilon_f
\end{eqnarray}
which tend to the isolated radiator frequency 
\begin{equation}
\omega_{\alpha} \equiv \omega_{fi} \equiv (E_f^{\infty}-E_i^{\infty})/\hbar
\end{equation}
as the perturbers get sufficiently far from the radiator:
\begin{eqnarray}
\omega_{e'e}(R) \rightarrow \omega_{fi} & \; {\rm as} 
& \; R \rightarrow \infty \;,   \; \; \; \; e  \; \in \varepsilon_i  \; ,  \; 
 e'  \; \in \varepsilon_f \; .
\end{eqnarray}
Let us introduce projectors ${\bf P}_e$ which select the $e^{th}$
adiabatic component of any $\Psi(x,R)$ according to~\cite{royer80}
\begin{equation}
{\bf P}_e\Psi(x,R) = \psi_e (R)\chi_e(x;R) \; .
\end{equation}
We write the dipole moment as a sum over transitions
\begin{equation}\label{eq:expdal}
{\bf D}  = \sum_{\alpha} {\bf D}_{\alpha} \; ,  
\end{equation}
\begin{equation}
{\bf D}_{\alpha} \equiv 
\sum_{e,e'} \, \! ^{(\alpha)}  \; {\bf P}_{e'} {\bf D} {\bf P}_{e} \; . 
\end{equation}
In the Heisenberg representation
\begin{eqnarray} 
{\bf D}_{\alpha}(t) & \equiv &  \sum_{e,e'} \, \! ^{(\alpha)}  \; 
      e^{\frac{i t {\bf H} }{\hbar }} {\bf P}_{e'}{\bf D}{\bf P}_e 
      e^{\frac{-i t{\bf H}}{\hbar }} \; , \\
      & \equiv &\sum_{e,e'} \, \! ^{(\alpha)}  \;  {\bf D}_{e'e}(t) \; .
\label{eq:dalpha}
\end{eqnarray}
The sum $\sum_{e,e'} ^{(\alpha)}$ is over all pairs ($e,e'$)  such that
\mbox{$\omega_{e',e}(R) \rightarrow \omega_{\alpha}$} as 
\mbox{$R \rightarrow \infty$}. 
Thus ${\bf D}_{\alpha}$ connects all pairs of adiabatic states whose
electronic energy differences become equal to  $\omega_{\alpha}$
as $R \rightarrow \infty$. In the absence of perturbers, 
${\bf D}_{\alpha}$ would be the component of ${\bf D}$ responsible for
the radiative transitions of frequency $\omega_{\alpha}$.
We note that the projection operators account for the 
weighting factors discussed in Ref.~\cite{allard94}. 

Introducing the expansion Eq.~(\ref{eq:expdal})
for ${\bf D}$ into the expression Eq.~(\ref{eq:autocorr}) for 
$\Phi(s)$, we obtain
\begin{equation}
\Phi(s) = \sum_{\alpha,\beta} \Phi_{\alpha,\beta}(s)
\end{equation} 
where
\begin{eqnarray}
\Phi_{\alpha,\beta}(s)
& = & {\bf Tr} \, \rho \, {\bf D}^{\dagger}_{\alpha} \, 
    e^{ \frac{i s {\bf H} }{\hbar }   } \,
      {\bf D}_{\beta} \,
    e^{ \frac{-i s {\bf H} }{\hbar }   }  \label{eq:phi_trace} \\
& = & \langle \,
       {\bf D}^{\dagger}_{\alpha}(0) \, {\bf D}_{\beta}(s) \,
      \rangle  \; .
\end{eqnarray}
The line shape is then
\begin{equation} \label{eq:shape1}
I(\omega) = \sum_{\alpha,\beta} I_{\alpha,\beta}(\omega) \; .
\end{equation} 
 The  terms  \mbox{$I_{\alpha,\beta}(\omega)$ }, \mbox{ $\alpha \neq \beta$},
represent interference between different spectral lines~\cite{baranger58b}.
When these interference terms are neglected, we get 
\begin{equation} \label{eq:shape2}
I(\omega) = \sum_{\alpha} I_{\alpha}(\omega)
\end{equation} 
and
\begin{equation} \label{eq:sumphi}
\Phi(s) = \sum_{\alpha} \Phi_{\alpha}(s)
\end{equation} 
where
\begin{equation}
\Phi_{\alpha}(s)=
\langle \, {\bf D}^{\dagger}_{\alpha}(0) \, {\bf D}_{\alpha}(s) \,
      \rangle  \; .
\end{equation} 
The time dependence of $\Phi_{\alpha}(s)$ is determined by 
${\bf D}_{\alpha}(s)$, the part of the dipole moment which,
in the absence of perturbers, oscillates at the frequency $ \omega_{\alpha} $.
Let us now denote 
\begin{equation}
{\bf  {d} }_{\alpha}(s)  \equiv  
{\bf D}_{\alpha} (s)e^{-i\omega_\alpha s}
\end{equation}
wherein the free evolution $e^{-i\omega_\alpha s}$ is factored out. 

For an isolated line, such as Lyman~$\alpha$, we have shown 
(Allard {\it et al.} \cite{allard99}) that the normalized
line shape $J_{\alpha}(\Delta \omega)$, in the uncorrelated perturbers
approximation, is given by
\begin{equation} 
  J_{\alpha}(\Delta \omega) = {\bf FT} [e^{ng_{\alpha}(s)}]
  \label{eq:J}
\end{equation}
In the classical path approximation, where we assume that the
 perturber  follows a rectilinear trajectory 
at a single mean velocity $\bar{v}$, we get 
from~\cite{allard94,allard99} 
that $g_{\alpha}(s)$ can be written as
\begin{eqnarray}
g_{\alpha}(s) && \,= \frac{1}
{\sum_{e,e'} \, \! ^{(\alpha)} \, |d_{ee'}|^2 }
\sum_{e,e'} \, \! ^{(\alpha)} \; \; \nonumber \\ 
&&  \int^{+\infty}_{0}\!\!2\pi\rho d\rho
\int^{+\infty}_{-\infty}\!\! dx \; 
\tilde{d}_{ee'}[ \, r(0) \, ] \, \nonumber \\ 
&&[ \, e^{\frac{i}{\hbar}\int^s_0 \, dt \;  
V_{e'e }[ \, r(t) \, ] } \,
\, \tilde{d^{*}}_{ee'}[ \, r(s) \, ] \, - \, \tilde{d}_{ee'}[ \, r(0) \, ] \, ]
\; .
\label{eq:final} 
\end{eqnarray}
The separation of the radiator and perturber is
\newline
 \mbox{ $r(t) = [ \, \rho ^2 + (x+\bar{v} t)^2 \, ]^{1/2}$ } with
$\rho$ the impact parameter of the perturber trajectory and 
$x$ is the position of the perturber along its trajectory at time
$t=0$. 
The total line strength of the transition is
$ \sum_{e,e'} \, \! ^{(\alpha)} \, |d _{ee'}|^2 $.
The potential energy for a state $e$ is 
\begin{equation}
V_{e}[r(t)] = E_e[ r(t) ]-E_e^{\infty} \; ;
\end{equation}
the difference potential is
\begin{equation}
V_{e' e}[r(t)] = V_{e' }[r(t)] - V_{ e}[r(t)] \; ;
\end{equation}
and we defined a {\em modulated} dipole \cite{allard99}  
\begin{equation}
\tilde{d}_{ee'}[r(t)] = 
d_{ee'}[r(t)]e^{-\frac{\beta}{2}V_{e}[r(t)] } \; ,
\end{equation}
where we denoted
\begin{equation}
d_{ee'}(   {\bf \vec{r}}   )=\langle \,\chi_e (  {\bf \vec{r}}   )|{\bf d}|
\chi_{e'} (  {\bf \vec{r}}   )\rangle \; . 
\end{equation}
In the above, we neglected the influence of the potentials
$V_e (r)$ and $V_{e'} (r)$ on the perturber trajectories, which remain
straight lines.
Although we should
drop the Boltzmann factor $e^{-\beta V_e (r)}$ for consistency with our
straight trajectory approximation, by keeping it we improve the
result in the wings. Note that over regions where $ V_e (r) <0 $, 
the factor $e^{-\beta V_e (r)}$ accounts for bound states of the
radiator-perturber pair, but in a classical approximation wherein the
discrete bound states are replaced by a continuum; thus any band
structure is smeared out.

\section{Theoretical analysis}
\label{sec:3}

\subsection{Formation of line satellites}
\label{satform}
Close collisions between a radiating  atom and a perturber
 are responsible for transient quasi-molecules 
 which may lead to the appearance of satellite features in
 the  wing of an atomic line profile.

When the difference $\Delta V(R)$ between the
upper and lower interatomic potentials for a given transition goes through an
extremum, a relatively wider range of interatomic distances contribute to the
same spectral frequency, resulting in an enhancement, or {\it satellite}, in
the line wing. The unified theory~\cite{anderson,allard82}
predicts that there will be satellites  centered periodically at 
frequencies corresponding to the extrema of the difference potential
between the upper and lower states, 
$\Delta\omega  =  k \Delta V_{\mathrm{ext}}, ( k  =  1,2, \ldots) $
\cite{allard78,royer78,kielkopf79}.
Here $\Delta\omega$ is the frequency difference between the center of
the unperturbed spectral line and the satellite feature, measured for
convenience in the same units as the potential energy difference.
This series of satellites is due to many-body interactions. 

\subsection {Diatomic potentials}
\label{pot}
The adiabatic interaction of the neutral hydrogen atom with a proton
or another hydrogen atom
is described by potential energies $V_e(R)$ for each electronic state of
the H$_2^+$ or H$_2$ molecule ($R$ denotes the internuclear distance between
the radiator and the perturber).
For H-H$^+$ collisions we have used the potentials of H$_2^+$ calculated 
by Madsen and Peek \cite{madsen71}. 
For H-H collisions we  have used the potentials of H$_2$ calculated 
by Schmelcher~\cite{schmelcher}.

\begin{figure*}
\resizebox{0.75\textwidth}{!}
{ \includegraphics{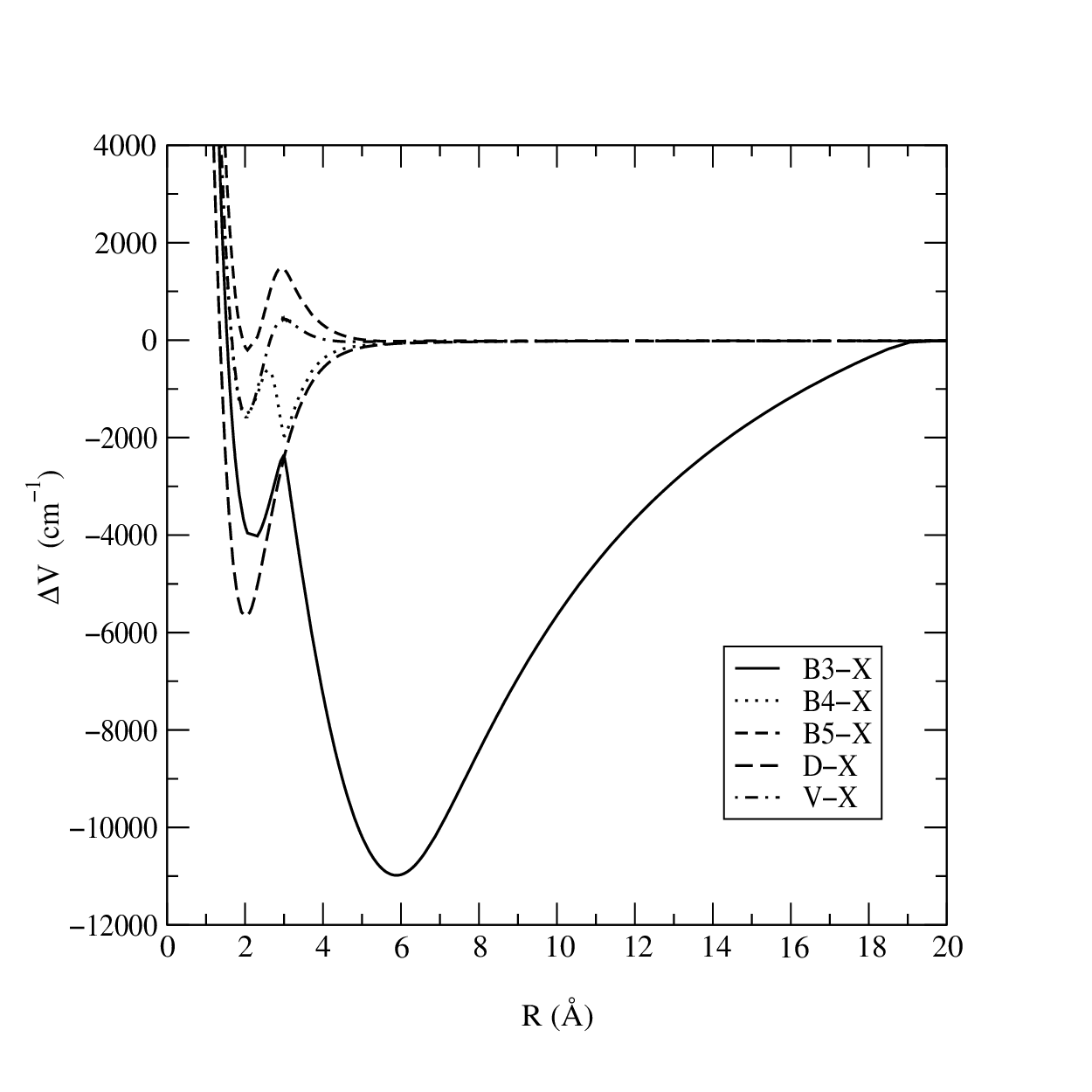} }
\caption{Difference potential energies expressed in
cm$ ^{-1}$ for the singlet states of H$_2$ contributing to Lyman~$beta$. }
\label{singlets}    
\end{figure*}

In Fig.~\ref{singlets} we have plotted the H$_2$ potential differences
 $\Delta V (R)$ for  the
singlet states which contribute to Lyman~$\beta$.  
We have used letters here to label the states.
B3 is the well known $\mathrm{B}''\bar{\mathrm{B}}\;^1\Sigma^+_u$ state. 
At small internuclear separation, $R$, the state correlates with the 
$\mathrm{B}''\;^1\Sigma^+_u $ state, the third $^1\Sigma^+_u $ state
of the Rydberg series. At larger internuclear separation the state has
 an ionic character until $R$= 19 \AA\/ where the potential energy curve  of
the $\mathrm{H}^+ + \mathrm{H}^- \; (1s)^2$  state crosses the
 $\mathrm{H} (n=1)  + \mathrm{H} (n=3)$ energy  levels, because of an
 avoid crossing $\bar{B}$ looses its  ionic character.
Dabrowski and Herzberg~\cite{dabrowski76}
 predicted the existence of the $\bar{B}$, the first calculations were
 done by Kolos~\cite{kolos76,kolos81}. More recent calculations 
are reported in \cite{detmer98,reinhold99}.

We use B4 and B5 respectively to label the {\hbox{4 $^1\Sigma^+_u$} and 
{\hbox{5 $^1\Sigma^+_u$} states, and
V and D to label the {\hbox{3 $^1\Pi_u$}} and {\hbox{2 $^1\Pi_u$}} states.

Each difference potential exhibits at least one extremum which,
in  principle,  leads to a corresponding
satellite  feature in the wing of Lyman~$\beta$ (see Sect.~\ref{satform}).
The present approach now allows us also to take into account
the  asymptotically forbidden transitions  of quasi-molecular
hydrogen which dissociate into  (1s,3s) and (1s,3d) atoms. 
The satellite amplitude depends on the value of the dipole moment  
through the  region of the potential extremum responsible of the
satellite and on the position of this extremum.
We have shown~\cite{allard99,allard98a,allard98b} that a large
enhancement in the amplitude of a spectral line satellite  occurs
whenever the   dipole moment  increases through  the region of
internuclear  distance where the  satellite is formed.

The  potential differences of the B3-X and B4-X transitions exhibit 
double wells. The maximum at 3.0~\AA\/ of the B3-X potential and
minimum of the B4-X potential are due to
an avoided-crossing. 

The most significant characteristic of Fig.~\ref{singlets} is the
existence of the deep  outer well at 6~\AA\/ of the B3-X potential.
The ionic interaction decays slowly making the potential energy difference
very broad compared to the very steep wells of the other transitions.
This is very important as the position of the extremum and the
functional  dependence of the  potential  difference on internuclear
separation determine the  amplitude  and shape of the
satellites~\cite{allard94}. 

\subsection{Electronic transition dipole moments}

The  dipole moment taken between the initial and
final states of a radiative transition determines the transition
probability, but for two atoms in collision, the moment depends
on their separation.  This modifies relative contributions to the
profile along the collision trajectory. 
Dipole moments for H$^+_2$ and  H$_2$, calculated as a function of 
internuclear distance respectively by Ramaker and Peek
\cite{ramaker72} and by Drira  \cite{drira99}, were used for
the  transitions contributing to Lyman~$\alpha$ and Lyman~$\beta$.
For Lyman~$\beta$, the four components which correspond to $1s-3p$ atomic
transitions are  dipole allowed \cite{drira99}: 
the two singlet B4-X and D-X transitions, and the two triplet
{4 \hbox{$^3\Sigma_g^{+}$}}-{\hbox{$^3\Sigma_u^{+}$}}
and  {\hbox{2 $^3\Pi_g$}}-{\hbox{$^3\Sigma_u^{+}$}} transitions .

If  electronic states $i$ and $f$ of an isolated radiator are not
connected by the dipole moment operator, that is if
$D_{if}(R \rightarrow \infty) = 0$, allowed radiative transitions
cannot occur between these two states. This  happens for
the other transitions which correspond to $1s-3s$ and $1s-3d$ atomic
transitions.
Although these transitions should not contribute to the unperturbed line
profile, $D_{if}(R)$ may differ
from zero when a perturber passes close to the radiator. In this instance
radiative transitions are induced by collisions, but not at the unperturbed
line frequency.

It often arises that an extremum in the potential difference
occurs when the final (or initial) potential energy curve exhibits an
avoided crossing, the corresponding wavefunctions exchange their
characteristics and the radiative dipole transition moment varies 
dramatically with $R$.
This is exactly what happens for the B3-X transition. 
To point out the importance of variation of dipole moment 
on the formation of  a collision-induced (CI) satellite, 
we have displayed in Fig.~\ref{forbidden} 
$D(R)$ together with the corresponding $\Delta V (R)$ for the $B3-X$
asymptotically forbidden transition. 
The dipole transition is extremely small for the isolated radiating
atom (\mbox{ $R \rightarrow \infty$ }) but it goes through a maximum
at the value of $R$ where the avoid-crossing occurs and
remains quite important at the internuclear distance where the
potential difference of the outer well goes through a minimum.

In such a case we expect a contribution from this transition in the
wing and the formation of a CI line satellite, if
it is not smeared out by larger dipole-allowed contributions. 

\begin{figure*}
\resizebox{0.75\textwidth}{!}
{ \includegraphics{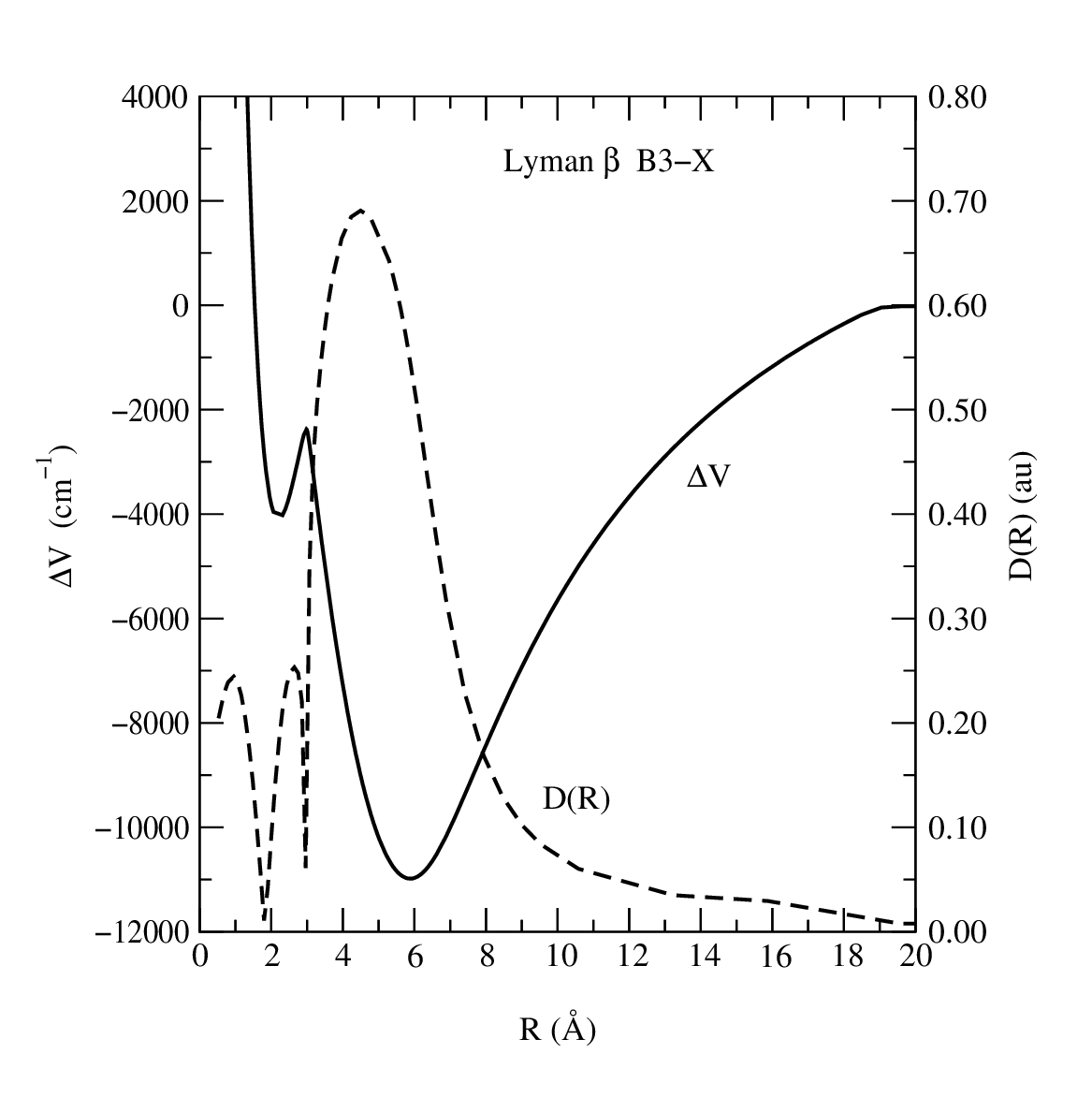} }
\caption{Difference potential energy $\Delta V$ in cm$^{-1}$ 
and the corresponding $D(R)$ in atomic units
for the dipole forbidden B3-X transition which gives the 
CI satellite of Lyman~$\beta$. }
\label{forbidden}
\end{figure*}

\section { Lyman~$\beta$ profile }

The Lyman profiles and satellites shown here are calculated at the low
densities met in the atmospheres of stars. The typical particle
densities from 10$^{15}$ to 10$^{17}$ cm$^{-3}$  allows
us to  use an expansion of the autocorrelation function in powers of
density  as described in~\cite{allard94,royer71}. Line profiles are 
normalized so that over $\omega$ they integrate to 1.

\subsection {Collisional profile perturbed by neutral H }
\label{lybh}
We will consider the two following mechanisms which contribute to the
Lyman~$\beta$ wing.
\begin{itemize}
\item The far wing of allowed dipole lines, due to the free-free
transition  in a pair of colliding atoms. 
\item The collision-induced absorption due to the free-free transition 
involving the transient dipole moment existing during a binary
collision. 
\end{itemize}
The line profile calculation shown Fig.~\ref{betatot}
has been done at a temperature of 10000~K for 
 a perturber density of $10^{16}$~cm$^{-3}$ of neutral hydrogen. 
The only line feature is a broad CI satellite
situated at 1150~\AA\/ in the far wing, due to the B3-X dipole
forbidden transition. 
Normally such an effect would be overshadowed by  the allowed transition wing, 
but in this case  there is no large contribution of the dipole
allowed transitions in this region,
as  can be easily predicted by the examination of 
Fig.~\ref{singlets}.
The extrema of the allowed B4-X and D-X
transitions occur for  very short distances, and are much smaller 
compared to the position  and depth of the outer well in the 
B3-X transition (see Sect.~\ref{satform} and ~\ref{pot}).

The collision-induced absorption depends on the internuclear
separation and produces very broad spectral lines with a
characteristic width of the order of the inverse
of  the duration of the close collision. 
It is strongly dependent on 
\begin{itemize}
\item the temperature
\item the amplitude of the dipole for  $R_{\rm min}$ when the
potential difference presents a minimum.
\end{itemize}
This emphasizes the importance
of  the accuracy of both the potential energies {\em and} the dipole moments
for the line shape calculations.  

\begin{figure*}
\resizebox{0.75\textwidth}{!}
{ \includegraphics{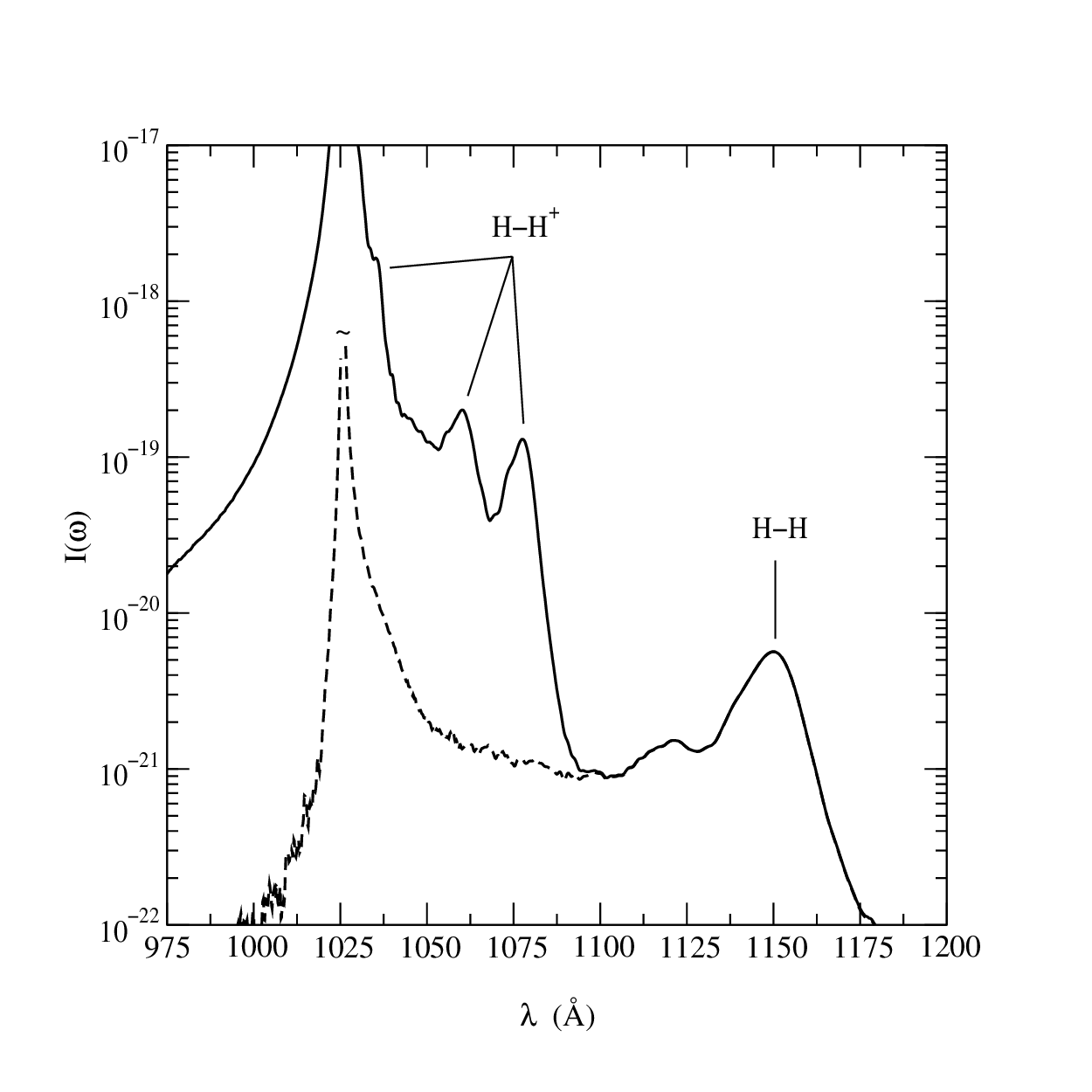} }
\caption{Line profile of Lyman~$\beta$ perturbed by neutral hydrogen
and protons. The dashed line ($---$) shows only the contribution from neutral
perturbers.}
\label{betatot}  
\end{figure*}

Oscillatory structures  appear near the 1150~\AA\/ satellite as they 
appear between the line and  the 1600~\AA\/ satellite in both 
theory and experiment~\cite{allard99}.  
These oscillations were predicted by Royer \cite{royer71b} and 
Sando {\it et al.}\/  \cite{sando69,sando73}. 

\subsection {Simultaneous perturbations by H and H$^+$}

The  complete Lyman~$\beta$  profile  perturbed by collisions with
neutral  hydrogen and protons is shown in Fig~\ref{betatot}.
We notice that the collision-induced H-H satellite 
is much broader than the allowed H-H$^+$ satellite since the dipole
moment differs from zero only over a  short range of internuclear
distances (see Sect.~\ref{lybh}). 
The CI  satellite of Lyman~$\beta$ is quite far from the 
unperturbed Lyman~$\beta$ line center,  actually closer
to the Lyman~$\alpha$ line. It is therefore necessary to take into proper
account the total
contribution of both the  Lyman~$\alpha$ and Lyman~$\beta$ wings of H
perturbed simultaneously by neutrals and protons and to study the 
variation  of this part of the Lyman series 
with the relative density of ionized and neutral atoms.

In~\cite{allard99} we  evaluated both the  Lyman~$\alpha$ and Lyman~$\beta$ wings of H
perturbed by protons.  However, we neglected interference terms
 between the  two lines.
 Equation~(51) of \cite{allard99}, which gives the profile for a pair of lines 
such as Lyman  $\alpha$ and  $\beta$, is
\begin{eqnarray}
I(\omega)= \phi_{\alpha}^{(0)} \, e^{nf_{\alpha}(0)} \, 
J_{\alpha}(\omega-\omega_{\alpha}) \nonumber \\ 
+ \,\phi_{\beta}^{(0)} \, e^{nf_{\beta}(0)} \, 
J_{\beta}(\omega-\omega_{\beta}) \; .
\end{eqnarray}
The perturbed line strength $\phi_{\alpha}^{(0)} e^{nf_{\alpha}(0)}$
differs from the {\it free}\/ line strength $\phi_{\alpha}^{(0)}$ by the
factor $e^{nf_{\alpha}(0)}$. This density-dependent factor
expresses the fact that the total intensity radiated increases or
decreases when the dipole moment is increased or decreased, on
average,  by the proximity of perturbers. 
Because of the low densities we consider, we have neglected this factor here. 

We show the sum of the profiles of Lyman~$\alpha$ and
Lyman~$\beta$ in Fig.~\ref{lytotal}.  We can see that a ratio of 5
between the neutral and proton density is
enough to make the CI satellite  appear in the far wing.
The CI satellite appearance is then very sensitive to the  degree of
ionization and may be used as a temperature diagnostic.

\begin{figure*}
\resizebox{0.75\textwidth}{!}
{ \includegraphics{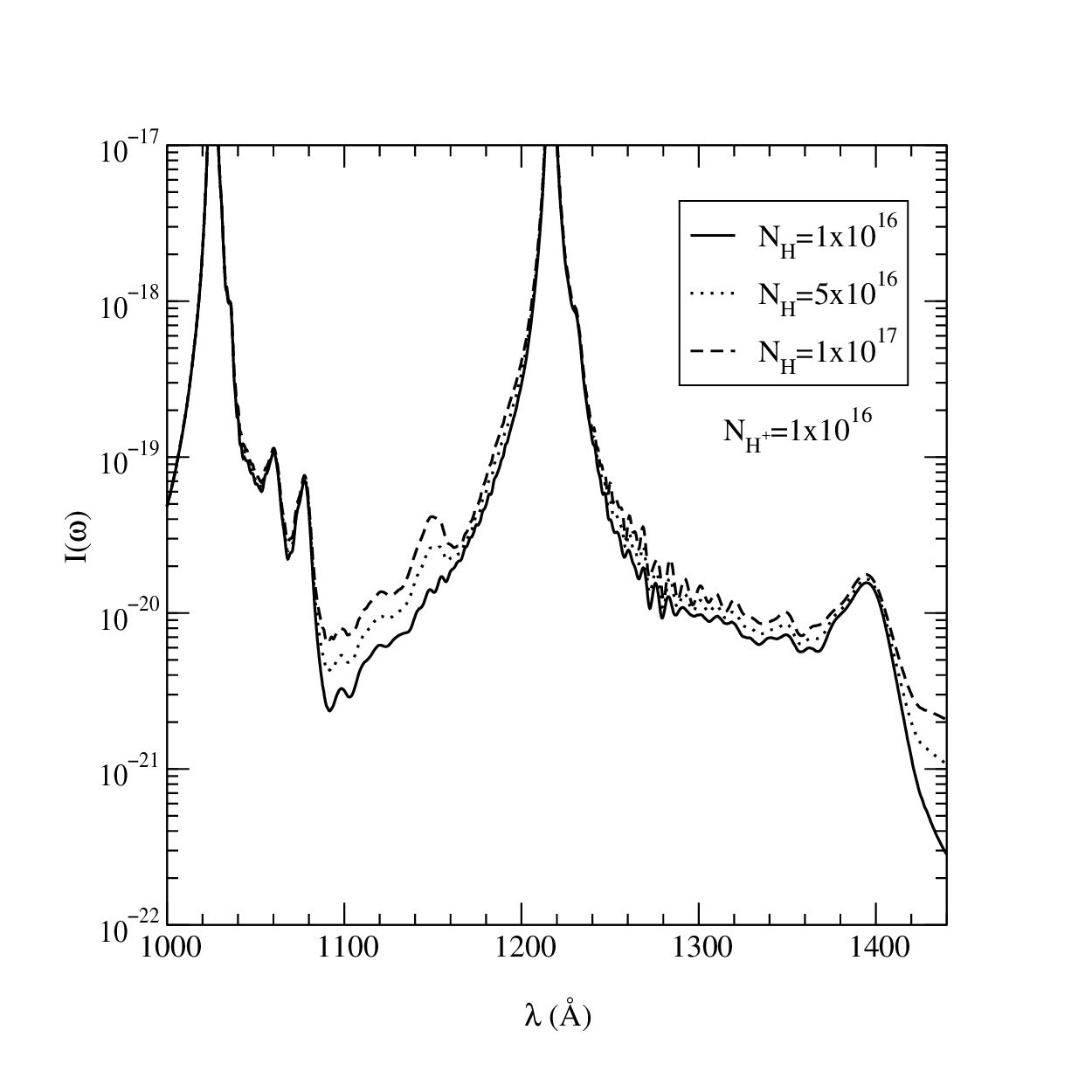} }
\caption{Total profile of Lyman~$\alpha$ and 
Lyman~$\beta$ perturbed by neutral hydrogen
and protons. Three different neutral densities ($1\times10^{16}$,
$5\times10^{16}$, $1\times10^{17}$ cm$^{-3}$) are compared for a fixed
ion density ($1\times10^{16}$ cm$^{-3}$).}
\label{lytotal} 
\end{figure*}

\section{Conclusions}

In the case of Lyman~$\alpha$ and the H-H$^+$ Lyman~$\beta$ satellites, 
the potential shape played a dominant role in the large difference 
in the broadening of the quasi-molecular features~\cite{allard98b}.
The width of the collision-induced absorption is determined, for the most
part, by the short range over which the corresponding 
transition dipole moment is significant.
In the CI satellite, it is the dipole moment which is very important
and which is responsible of the shape  of the satellite, the
observation of such a satellite would be a test of 
the accuracy of the dipole moment calculation. 
Satellites due to allowed and forbidden transitions 
depend linearly on density. The CI satellite is very sensitive to
the temperature of the absorption, and it may also be used
as a diagnostic  tool for temperature.
We emphasize that the effect of finite collision
duration does play a role in the shape of  the far wing. 
The present calculations are done in an adiabatic approximation
using a rectilinear trajectory.  This should affect slightly the
shape of the satellite, although no great error is expected.  We are 
developing methods to include trajectory effects in the evaluation of the
line shape.

\section*{Acknowledgements}

The computations of  dipole transition moments were 
performed on the CRAY of the computer center IDRIS.
The work at the University of Louisville is supported by a grant
from the U.S. Department of Energy, Division of Chemical Sciences,
Office of Basic Energy Sciences, Office of Energy Research.

%

%

\begin{thebibliography}{}
%
%
%
\bibitem{allard99} 
N.F. Allard, A. Royer, J.F. Kielkopf, N. Feautrier, Phys. Rev. A
\textbf{60}, 1021 (1999).
\bibitem{anderson} 
P.W. Anderson, J.D. Talman , Bell System Technical 
Publication No. 3117 (Murray Hill, NJ, 29 1956).
\bibitem{baranger58a} 
M. Baranger, Phys. Rev. \textbf{111}, 481 (1958).
\bibitem{baranger58b} 
 M. Baranger, Phys. Rev. \textbf{111}, 494 (1958).
\bibitem{koester93} 
D. Koester, N.F. Allard, in  \textit{ White Dwarfs: Advances in 
Observation and Theory}, edited by M. Barstow (Kluwer, Dordrecht,1993),
p. 237.
\bibitem{koester94} 
D. Koester, N.F. Allard, G. Vauclair, Astron. Astrophys. \textbf{291}, 
L9 (1994).
\bibitem{holweger94}
H. Holweger, D. Koester, N.F. Allard, Astron. Astrophys.
\textbf{290}, L21 (1994)
\bibitem{bergeron95}
P. Bergeron, F. Wesemael, R. Lamontagne, G. Fontaine, R. Saffer, 
N.F. Allard, Astrophys. J. \textbf{449}, 258 (1995).
\bibitem{kielkopf95} 
J.F. Kielkopf, N.F. Allard, Astrophys. J. \textbf{450},  L75 (1995).
\bibitem{kielkopf98} 
J.F. Kielkopf, N.F. Allard, Phys. Rev. A \textbf{58},  4416 (1998).
\bibitem{koester96}
D. Koester, D.S. Finley, N.F. Allard, J.W. Kruk, R.A. Kimble,
Astrophys. J. \textbf{463}, L93 (1996).
\bibitem{koester98}
D. Koester, U. Sperhake, N.F. Allard, D.S. Finley, S. Jordan,  
Astron. Astrophys.  \textbf{336}, 276 (1998).
\bibitem{allard98a}
N.F. Allard, J.F. Kielkopf, N. Feautrier, Astron. Astrophy.
\textbf{330}, 782 (1998).
\bibitem{madsen71} 
Madsen M.M., Peek J.M.,  Atomic Data \textbf{2}, 1971, 171.
\bibitem{ramaker72} 
Ramaker D.E., Peek J.M.,  J. Phys. B \textbf{5}, 1972, 2175.
\bibitem{drira99}
Drira, I., J. Mol. Spectroscopy \textbf{198}, 52 (1999). 
\bibitem{schmelcher}
Schmelcher, P., private communication.
\bibitem{detmer98}
T. Detmer, P. Schmelcher, L.S. Cederbaum, J. Chem. Phys. \textbf {109}, 
9694 (1998).
\bibitem{royer74}  
A. Royer, Can. J. Phys. \textbf{52}, 1816 (1974).
\bibitem{royer80}  
A. Royer, Phys. Rev. A \textbf{22}, 1625 (1980).
\bibitem{allard94}
N.F. Allard, D. Koester, Feautrier N., Spielfiedel A., 
Astron. Astrophy. Suppl. Ser. \textbf{108}, 417  (1994).
\bibitem{allard82}
N.F. Allard, J. Kielkopf, Rev.~Mod.~Phys. \textbf{54}, 1103 (1982).
\bibitem{allard78}
N.F. Allard, J. Phys. B \textbf{11}, 1383 (1978).
\bibitem{royer78}  
A. Royer, Acta~Phys.~Pol.~A \textbf{54}, 805 (1978).
\bibitem{kielkopf79}
J.F. Kielkopf, N.F. Allard, Phys. Rev. Lett. \textbf {43}, 3 (1979).  
\bibitem{dabrowski76}
I. Dabrowski, G. Herzberg, Can. J. Phys. \textbf {54}, 525 (1976).
\bibitem{kolos76}
W. Kolos, J. Mol. Spectroscopy \textbf{62}, 429 (1976).
\bibitem{kolos81}
W. Kolos, J. Mol. Spectroscopy \textbf{86}, 420 (1981).
\bibitem{reinhold99}
Reinhold E., Hogervorst W., Ubachs W., Wolniewicz L. , Phys. Rev. A 
\textbf{60}, 1258 (1999)
\bibitem{allard98b}
N.F. Allard, I. Drira, M. Gerbaldi, J.F. Kielkopf, A. Spielfiedel,
Astron.  Astophys.  \textbf{335}, 1124 (1998).
\bibitem{royer71}
A. Royer, Phys. Rev. A \textbf{3}, 2044 (1971).  
\bibitem{royer71b}
A. Royer, Phys. Rev. A, \textbf{43}, 499 (1971). 
\bibitem{sando69} 
Sando K.M., Doyle R.O., Dalgarno A., Astrophys. J. \textbf{157}, L143 (1969). 
\bibitem{sando73}
Sando K.M., Wormhoudt J.G., Phys. Rev. A  \textbf{7}, 1889 (1973). 

\end{thebibliography}
%

\end{document}